\documentstyle[eqsecnum,aps]{revtex}
\begin{document}
\newcommand{\reals}{\mbox{${\rm I\!R }$}}
\newcommand{\komplex}{\mbox{${\rm I\!\!\!C }$}}
\newcommand{\nats}{\mbox{${\rm I\!N }$}}
\newcommand{\intgs}{\mbox{${\rm Z\!\!Z }$}}
\newcommand{\cab}{{\cal B}}
\newcommand{\can}{{\cal N}}
\newcommand{\cam}{{\cal M}}
\newcommand{\caz}{{\cal Z}}
\newcommand{\cao}{{\cal O}}
\newcommand{\cac}{{\cal C}}
\newcommand{\cah}{{\cal H}}
\newcommand{\al}{\alpha}
\newcommand{\be}{\beta}
\newcommand{\de}{\delta}
\newcommand{\ep}{\epsilon}
\newcommand{\ga}{\gamma}
\newcommand{\la}{\lambda}
\newcommand{\om}{\omega}
\newcommand{\ze}{\zeta}
\newcommand{\De}{\Delta}
\newcommand{\Ga}{\Gamma}
\newcommand{\Om}{\Omega}
\newcommand{\Si}{\Sigma}
\newcommand{\rS}{{\rm S}}
\newcommand{\snnu}{\sum_{n=0}^{\infty}}
\newcommand{\slnu}{\sum_{l=0}^{\infty}}
\newcommand{\snuu}{\sum_{n=-\infty}^{\infty}}
\newcommand{\sluu}{\sum_{l=-\infty}^{\infty}}
\newcommand{\sied}{\sum_{i=1}^3}
\newcommand{\sneu}{\sum_{n=1}^{\infty}}
\newcommand{\pxi}{\prod_{i=1}^3\left(1-e^{-x_i n}\right)}
\newcommand{\pxj}{\prod_{j=1}^3\left(1-e^{-x_ j n }\right)}
\newcommand{\en}{e^{-n\ep\al}}
\newcommand{\enx}{e^{-n\ep\al-nx_i}}
\newcommand{\exi}{(1-e^{-nx_i})}
\newcommand{\exj}{(1-e^{-nx_j})}
\newcommand{\xij}{x_1x_2+x_1x_3+x_2x_3}
\newcommand{\xp}{x_1x_2x_3}
\newcommand{\zrz}{\zeta_R (2)}
\newcommand{\zrd}{\zeta_R (3)}
\newcommand{\zrv}{\zeta_R (4)}
\newcommand{\ent}{d (\nu )}
\newcommand{\pnenj}{\prod_{j=1}^p\nu_j}
\newcommand{\nn}{\nonumber}
\renewcommand{\theequation}{\mbox{\arabic{equation}}}
\newcommand{\intsi}{\int\limits_{\Sigma}d\sigma_x\,\,}
\newcommand{\back}{\bar{\Phi}}
\newcommand{\coba}{\bar{\Phi}^{\dagger}}
\newcommand{\abl}{\partial}
\newcommand{\pa}{\partial}
\newcommand{\qpi}{(4\pi)^{\frac{q+1} 2}}
\newcommand{\snenp}{\sum_{n_1,...,n_p=0}^{\infty}}
\newcommand{\tint}{\int\limits_0^{\infty}dt\,\,}
\def\beq{\begin{eqnarray}}
\def\eeq{\end{eqnarray}}
\newcommand{\zb}{\zeta_{{\cal B}}(}
\newcommand{\rzb}{Res\,\,\zb}
\newcommand{\zn}{\zeta_{{\cal N}}\left(} 
\newcommand{\rzn}{Res\,\,\zn}
\newcommand{\fr}{\frac}
\newcommand{\sip}{\frac{\sin (\pi s)}{\pi}}
\newcommand{\rzs}{R^{2s}}
\newcommand{\g}{\Gamma\left(}
\newcommand{\ikma}{\int\limits_{\ga}\frac{dk}{2\pi i}k^{-2s}\frac{\pa}{\pa k}}
\newcommand{\suani}{\sum_{a=0}^i}
\newcommand{\zem}{\zeta_{{\cal M}}}
\newcommand{\hem}{A^{\cam}}
\newcommand{\hen}{A^{\can}}
\newcommand{\man}{{\cal M}}
\newcommand{\pold}{D^{(d-1)}}
\newcommand{\zesd}{\zeta_{S^d}}
\newcommand{\fac}{\frac{(4\pi)^{D/2}}{a^d|S^d|}}
\newcommand{\sri}{\sum_{i=1}^ d  r_i}
\newcommand{\pri}{\prod_{i=1}^d r_i}
\newcommand{\ber}{B^{(d)}}
\newcommand{\ar}{a|\vec r )}

\title{The $a_{5}$ heat kernel coefficient on a manifold with boundary} 
\author{
Klaus Kirsten$^{1}$
\thanks{Electronic address: klaus.kirsten@itp.uni-leipzig.de}}
\address{
${ }^{1}$Universit\"{a}t Leipzig, 
Institut f\"{u}r Theoretische Physik,\\
Augustusplatz 10, 04109 Leipzig, Germany}
\maketitle
\begin{abstract}
In this letter we present the calculation of the $a_{5}$ heat kernel
coefficient 
of the heat operator trace for a partial differential operator of 
Laplace type on a compact Riemannian manifold with Dirichlet and 
Robin boundary conditions.
\end{abstract}
\vspace{1cm}
Motivated by the need to give answers to some fundamental questions in quantum 
field theory, during the last years there has been and continues to be 
a lot of interest in the problem of calculating the heat kernel 
coefficients of partial differential operators (mostly of Laplace type) 
(see, for example \cite{1,2,3,3a,3b}). The coefficients contain 
information about the scaling and divergence behaviour of the 
quantum field theory. Furthermore, they determine effective actions 
at high temperature \cite{T1,T2,T3} or for large masses of the fields
involved 
\cite{mass} and they  
provide expansions in different background fields \cite{5}. 
In mathematics the interest in the 
heat kernel coefficients stems, in particular, from the well known 
connection that exists between the heat equation and the Atiyah-Singer 
index theorem \cite{4}. 

In general, if the manifold ${\cal M}$ (which we assume to be a compact 
Riemannian one) has a boundary $\partial {\cal M}$, the coefficients 
$a_n$ in the short-time expansion have a volume and a boundary part. For the 
volume part very effective systematic schemes have been developed 
(see for example \cite{5,6,7}). The calculation of the boundary part 
is in general more difficult. Only quite recently has the coefficient 
$a_4$ for Dirichlet and Robin boundary conditions been found \cite{8,9} 
(see also \cite{10,11,12,13,14,15,16}). 

For the $a_{5}$ coefficient only partial results are available 
\cite{17}. To state them in detail let us introduce some notation 
(see also \cite{17}, but with characteristic differences).
Let $V$ be a smooth vector bundle over $\cam$ equipped with a connection
$\nabla ^V$ and $E$ an endomorphism of $V$. Define the partial differential
operator 
\beq
D= -g^{ij} \nabla_i^V \nabla _j^V - E .\label{1}
\eeq
Furthermore we must impose suitable boundary conditions. If 
$\phi \in C^{\infty} (V)$, let $\phi_{;m} $ be the covariant derivative with 
respect to the exterior unit normal, let $S$ be an endomorphism of 
$V|_{\partial \cam}$. We define the Dirichlet boundary operator 
$\cab ^-$ and the Robin boundary operator $B_S^+$ by 
\beq
\cab ^- \phi \equiv \phi|_{\partial \cam } \quad 
\mbox{and} \quad \cab_S^+ \phi \equiv \left( \phi_{;m} -S\phi \right) 
|_{\partial \cam}   . \label{2} 
\eeq
To have a uniform notation we set $S=0$ for Dirichlet boundary conditions
and write $B_S^\mp$. Let $D_\cab$ be the operator defined by the appropriate 
boundary conditions. If $F$ is a smooth function on $\cam$, there is an 
asymptotic series as $t \to 0$ of the form 
\beq
\mbox{Tr} _{L^2} \left( F e^{-tD_\cab} \right) \approx 
\sum_{n\geq 0 } t^{\frac{n-m} 2 } a_n (F,D,\cab) , \label{3}
\eeq
where the $a_n (F,D,\cab)$ are locally computable \cite{18} .

We are now ready to state the results of \cite{17}. 
The coefficient $a_5$ for an arbitrary smearing function $F$ has been 
calculated for totally geodesic boundary $\partial\cam$. 
Furthermore, putting 
$F=1$ it has been found for $\cam$ being a domain in $\reals ^m$. Finally
it has been shown that for a smooth but not necessarily totally geodesic 
boundary, there exist universal constants so that
\beq
\lefteqn{
a_{5} (F,D,B_S^{\mp})=\mp 5760^{-1}(4\pi)^{-(m-1)/2}{\rm Tr} 
\{F\left\{g_1 E_{;mm} +g_2 E_{;m} S + g_3 E^2 \right. } \nn\\
& &+g_4  E_{:a}^{\phantom{:a}a} 
+ g_5 RE + 120 \Omega_{ab} \Omega^{ab} + g_6 \Delta R +
g_7 R^2 +g_8 R_{ij} R^{ij} +g_9  R_{ijkl} R^{ijkl} \nn\\
& &+g_{10} R_{mm} E 
+g_{11} R_{mm} R +g_{12} RS^2 +(-360^-,90^+)\Omega_{am} 
\Omega^a_{\phantom{a} m} +g_{13} R_{;mm} 
+g_{14} R_{mm:a}^{\phantom{mm:a}a} \nn\\
& &+g_{15} R_{mm;mm} 
+g_{16} R_{;m} S+g_{17} R_{mm} S^2 
+g_{18} S S_{:a}^{\phantom{:a}a} +g_{19} S_{:a} S_:^a 
+g_{20} R_{ammb} R^{ab} \nn\\
& &\left.+g_{21} R_{mm} R_{mm} 
+g_{22} R_{ammb} R^{a\phantom{mm}b}_{\phantom{a}mm} +
g_{23} ES^2 +g_{24} S^4\right\} \nn\\
& &+F_{;m} \left\{g_{25} R_{;m} +g_{26}RS +g_{27} R_{mm} S 
+g_{28} S_{:a} ^{\phantom{:a}a} 
+g_{29} E_{;m} +g_{30} ES +g_{31} S^3\right\}\nn\\
& &+F_{;mm} \left\{g_{32} R +g_{33} R_{mm} +g_{34} E +g_{35} S^2 \right\}
+g_{36}  S F_{;mmm} +g_{37} F_{;mmmm}\nn\\
& &+F\left\{d_1 KE_{;m} +d_2 KR_{;m} +
d_3 K^{ab} R_{ammb;m} +d_4 K S_{:b}^{\phantom{:b}b} 
+d_5 K_{ab} S_:^{ab} \right.\nn\\
& &+d_6 K_{:b} S_:^b +d_7 {K_{ab:}}^a S_:^b
+d_8 {K_{:b}}^b S +d_9 {K_{ab:}}^{ab} S 
+ d_{10} K_{:b} K_:^b 
+d_{11} {K_{ab:}}^a K_:^b \nn\\
& &+d_{12} {K_{ab:}}^a {K^{bc}}_{:c}
+d_{13} K_{ab:c} K^{ab\phantom{:}c}_{\phantom{ab}:}
+d_{14} K_{ab:c} K^{ac\phantom{:}b}_{\phantom{ac}:} 
+d_{15} K_{:b}^{\phantom{:b}b} K \nn\\ 
& &+d_{16} K_{ab:}^{\phantom{ab:}ab} K 
+d_{17} K_{ab:\phantom{a}c}^{\phantom{ab:}a} K^{bc} 
+d_{18} K_{:bc}K^{bc} +d_{19} K_{bc:a}^{\phantom{bc:a}a} 
K^{bc}\nn\\ 
& &+g_{38} KSE +d_{20} KS R_{mm} +g_{39} KSR 
+d_{21} K_{ab} R^{ab} S +d_{22} K^{ab} S R_{ammb}
+g_{40} K^2 E \nn\\
& &+g_{41} K_{ab} K^{ab} E 
+g_{42} K^2 R
+g_{43} K_{ab}K^{ab} R +d_{23} K^2 R_{mm} \nn\\
& &+d_{24} K_{ab} K^{ab} R_{mm} +d_{25} KK_{ab} R^{ab} 
+d_{26} KK^{ab} R_{ammb} +d_{27}K_{ab}K^{ac}R^b_c \nn\\
& &+d_{28} K_a^b K^{ac} R_{bmmc} +d_{29} K_{ab} K_{cd} R^{acbd} +
d_{30} KS^3 +d_{31} K^2 S^2 +d_{32} K_{ab} K^{ab} S^2 \nn\\
& &+d_{33} K^3 S 
+d_{34} KK_{ab} K^{ab} S +d_{35} K_{ab} K^{bc} K^a_c S + 
d_{36} K^4 +d_{37} K^2 K_{ab} K^{ab} \nn\\
& &\left.+d_{38} K_{ab}K^{ab} K_{cd} K^{cd} + d_{39} 
KK_{ab} K^{bc} K_c^a + d_{40} K_{ab} K^{bc} K_{cd} K^{da} \right\} 
\nn\\
& &+F_{;m} \left\{g_{44} KE +d_{41} KR_{mm} +
g_{45} KR+d_{42} KS^2 
\right.\nn\\
& &+d_{43} K_{:b}^{\phantom{:b}b} + 
d_{44} K_{ab:}^{\phantom{ab:}ab} 
+d_{45} K_{ab}R^{ab} +d_{46} K^{ab} R_{ammb} 
+d_{47} K^2S \nn\\
& &\left.+d_{48}K_{ab} K^{ab} S +d_{49} K^3 
+d_{50} KK_{ab} K^{ab} +d_{51} K_{ab} K^{bc} K^a_c
\right\}\nn\\
& &+F_{;mm} \left\{d_{52} KS +d_{53} K^2 +d_{54} 
K_{ab} K^{ab} \right\} +d_{55} K  F_{;mmm}\} [\partial {\cal M}]
\label{4}
\eeq
Here and in the following $f[\cam]=\int_{\cam}dx\, f(x) $ 
and $f[\partial \cam] = \int_{\partial \cam} dy f(y) $,
with $dx$ and $dy$ being the Riemannian volume elements 
of $\cam$ and $\partial \cam$. In addition, "$;$" denotes differentiation 
with respect to the Levi-Civita connection of $\cam$ and "$:$" 
covariant differentiation tangentially with respect to the Levi-Civita 
connection of the boundary. Finally, our sign convention is 
$R^i_{\phantom{i}jkl} = -\Gamma^i_{jk,l}+\Gamma^i_{jl,k} +
           \Gamma^i_{nk} \Gamma^n_{jl} -
           \Gamma^i_{nl} \Gamma^n_{jk}
          $ (see for example \cite{haw}). 

In addition to the terms containing the curvature $\Omega$ of 
the connection $\nabla ^V$,  in \cite{17} the values of 
$g_1,...,g_{45}$ and $d_{43},d_{44}$ and $d_{55}$ were calculated.
However, the main group of terms containing the extrinsic 
curvature $K_{ab}$ and its derivatives remained undetermined and
will be found here for the first time. In order to explain the way
we proceed we will take $g_1,...,g_{45}$ and $d_1,...,d_{55}$ as unknown.
The small group containing the curvature $\Omega$ is completely known and 
will have no influence on the calculation presented here.

We are going to use essentially three different ingredients. The first 
one is a Lemma on product manifolds \cite{8}. 
Let $N^{\nu} (F) = F_{;m...}$ be the $\nu^{th}$ normal covariant 
derivative.
There exist local
formulae $a_n (x,D)$ and $a_{n,\nu} (y,D)$ so that 
\beq
a_n (F,D,\cab _S^\mp ) = \{Fa_n (x,D)\} [\cam] +
 \{  \sum_{\nu = 0}^{2n-1} N^\nu (F) a_{n,\nu} (y,D,\cab_S^\mp )\}
   [\partial \cam].
\nn
\eeq
Let $\cam = \cam _1 \times \cam _2$ and $D=D_1 \otimes 1 + 1\otimes D_2$ 
and $\partial \cam _2  = \emptyset $. Then 
\beq
a_{n,\nu} (y,D,\cab _S ^\mp ) = \sum_{p+q=n} a_{p,\nu} (y_1,D_1,\cab _S^\mp )
a_q (x_2,D_2).\nn
\eeq
For $a_5$ this means 
\beq
a_5 (y,D,\cab_S^\mp ) - a_5 (y_1,D_1,\cab _S ^\mp ) a_0 (x_2,D_2) = 
a_3 (y_1,D_1,\cab_S^\mp ) a_2 (x_2,D_2) +a_1(y_1,D_1,\cab_S^\mp ) 
      a_4 (x_2,D_2).
\label{5}
\eeq

This gives the following $22$ universal constants: 
\begin{equation}
\begin{array}{llll}
g_3 = 720 & g_5 = 240 & 
g_6 = 48 & g_7 = 20 \\
 g_8= -8 & g_9 = 8 & 
g_{10} = -120 & g_{11} 
=-20 \\
 g_{12} = 480 & g_{23} = 2880 & g_{26} = -240 & g_{30} -1440 \\ 
g_{32} = 60 & g_{34} = 360 & g_{38} = 1440 & g_{39} = 240 \\ 
g_{40} = (105^-, 195^+) & g_{41} = (-150^-, 30^+) & 
g_{42} = 
(105/6^-, 195/6^+) & g_{43} = (-25^-,5^+) \\
 g_{44}= (450^-,-90^+) &
g_{45} = (75^-,-15^+) & & 
\end{array}
\end{equation}
Next we use the calculations on the $m$-dimensional ball. For $F=1$ 
the results are given in \cite{19,20}, the generalization to arbitrary
$F$ has been achieved recently and will be presented in \cite{21}.
This leads to the following additional $25$ constants or 
relations among them: 
\beq
\begin{array}{llll}
g_{24} = 1440 & g_{31} = -720 & g_{35} = 360 \\
g_{36} = -180 &
g_{37} = 45 & d_{30} = 2160 \\
 d_{31} = 1080 &
d_{32} = 360 & 
d_{33} = 885/4 \\
 d_{34} = 315/2 &
d_{35} = 150 & d_{36} = (-65/128^-, 2041/128^+)  \\
d_{37} = (-141/32^-,417/32^+) &
d_{40} = (-327/8^-, 231/8^+) &
d_{42} = -600 \\
 d_{47} = -705/4 & 
d_{48} = 75/2 & d_{49} 
= (495/32^-, -459/32^+) \\
 d_{50} = (-1485/16^-, -267/16^+) & 
d_{51} = (225/2^-, 54^+) & 
d_{52} = 30 \\
d_{53} = (1215/16^-, 315/16^+) &d_{54} = (-945/8^-, -645/8^+) &
d_{55} = (105^-,30^+) 
\end{array}
\eeq
and $d_{38} +d_{39} =(1049/32^-, 1175/32^+)$.

All this information is now a very good starting point to use relations 
of the heat kernel coefficients under conformal rescalings \cite{8}.
The relevant ones for our case read
\beq
\frac d {d\ep} |_{\ep =0} a_5 \left( 1, e^{-2\ep F} D\right))- 
(m-5) a_5 (F,D) &=& 0\label{8}\\
\frac d {d\ep} |_{\ep =0} a_5 \left( e^{-2\ep f} F, e^{-2\ep f} D 
\right) &=& 0 
\quad \mbox{for } m=7.\label{9}
\eeq
Setting to zero the coefficients of all terms in (\ref{8})
we obtain the following equations. (They are ordered in a way such that
nearly every equation immediately yields a universal constant. This 
was the main motivation for the given ordering.) 
\beq
\begin{array}{ll}
\underline{\mbox{Term}} & \underline{\mbox{Coefficient}} \\
EF_{;mm} & 0=-2g_1 +(m-2)g_3 -2(m-1)g_5 -(m-1)g_{10} - (m-5) g_{34} \\
ESF_{;m} & 0=-2g_2 -(m-2)g_{23} +(m-1) g_{38} -(m-5) g_{30} \\
SF_{;mmm} & 0=\frac 1 2 (m-2) g_2 -2(m-1)g_{16} -(m-5) g_{36}  \\
KSF_{;mm} & 0=\frac 1 2 (m-2) g_2 -2(m-1) g_{16} +\frac 1 2 (m-2) g_{38} 
         -(m-1) d_{20} -2(m-1) g_{39} -d_{21} +d_{22} -(m-5) d_{52} \\
FE_{:a}^{\phantom{:a} a} & 0=-g_1 +(m-2) g_3 -(m-5) g_4 -2(m-1) g_5 -g_{10} \\
F_{;mmmm} & 0=\frac 1 2 (m-2) g_1 -2(m-1) g_6 -2(m-1) g_{13} -(m-1) g_{15} 
            -(m-5) g_{37} \\
F\Delta R & 0=\frac 1 2 (m-2) g_5 -(m-4) g_6 -4(m-1) g_7 -mg_8 -4g_9 -g_{11}
           -g_{13} +\frac 1 2 g_{20} \\
FR_{;mm} & 0= -\frac 1 2 (m-2) g_5 +(m-4) g_6 +4(m-1) g_7 +2(m-1) g_8 +8g_9 
            +g_{11} +g_{13} -2g_{15} -\frac m 2 g_{20} +g_{22} \\
FR_{mm:a}^{\phantom{mm:a} a} & 0=\frac 1 2 (m-2) g_1 -2(m-1) g_6 +\frac 1 2 
       (m-2) g_{10} -2(m-1) g_{11} -2(m-1) g_{13} -(m-5) g_{14} -2g_{15} \\
       & \phantom{0=}+(m-1) 
        g_{20} -2g_{21} -2g_{22} \\
F_{;mm} S^2 & 0=-2(m-1) g_{12} -(m-1) g_{17} +\frac 1 2 (m-2) g_{23} 
            -(m-5) g_{35} \\
FS_{:a} S_:^a & 0= -4(m-1) g_{12} -2g_{17} -(m-3) g_{18} +2g_{19} 
                +(m-2) g_{23} \\
F_{;m} E_{;m} & 0= -5g_1 -\frac 1 2 (m-2) g_2 +(m-1) d_1 -(m-5) g_{29} \\  
F_{;mmm} K & 0= \frac 1 2 (m-2) g_1 -4(m-1) g_6 -2(m-1) g_{13} -g_{15} 
        +\frac 1 2 (m-2) d_1 -2(m-1) d_2 +d_3 -(m-5) d_{55} \\
F_{;m} R_{;m} & 0= -\frac 1 4 (m-2) g_1 +(2m-7) g_6 +(m-6) g_{13} -2g_{15} 
          -\frac 1 2 (m-2) g_{16} +(m-1) d_2 -\frac 1 2 d_3 -(m-5) g_{25} \\
F_{;mm} R_{mm} & 0= -(m-2) g_1 +4(m-1) g_6 -2(m-2) g_8 -8g_9 +\frac 1 2
          (m-2) g_{10} -2(m-1) g_{11} \\
           & \phantom{0=} +4(m-1) g_{13} -2(m-1) g_{21} -2g_{22}
               -(m-5) g_{33}   \\
F_{;m} R_{mm} S & 0= -\frac 1 2 (m-2) g_2 +2(m-1) g_{16} -(m-2) g_{17} 
          +(m-1) d_{20} -d_{21} -d_{22} -(m-5) g_{27} \\
FKS_{:a}^{\phantom{:a} a} & 0= -(m-4) d_4 -d_5 +d_6 +d_7 -d_8 -d_9 +\frac 
             1 2 (m-2) g_{38} -d_{20} -2(m-1) g_{39} -d_{21} \\
FK_{:a}^{\phantom{:a} a} S & 0= -\frac 1 2 (m-2) g_2 +2(m-1) g_{16} 
                   -d_4 +d_6 -(m-4) d_8 +\frac 1 2 (m-2) g_{38} -d_{20}
           -2(m-1) g_{39} -d_{21} \\
FK_{ab} S_:^{ab} & 0= -(m-2) g_2 +4(m-1) g_{16} +3d_5 -(m-2) d_7 
                +(m-2) d_9 -(m-2) d_{21} +d_{22} \\
FK_{ab:}^{\phantom{ab:} ab} S & 0= -d_5 +d_7 -(m-4) d_9 -(m-2) d_{21} +d_{22} \\
F_{;m} S_{:a} ^{\phantom{:a} a} & 0= \frac 1 2 (m-2) g_2 -2(m-1) g_{16} -(m-2)
             g_{18} +(m-2) g_{19} +(m-1) d_4 +d_5 \\
                    & \phantom{0=}  -(m-1) d_6 -d_7 
             +(m-1) d_8 +d_9 -(m-5) g_{28} 
\end{array}
\eeq
The equations given up to this point allow for the determination 
of the universal constants apart from two groups. The first group 
is $d_{23},...,d_{29},d_{38},d_{39}, d_{41}, d_{45}, d_{46}$. The second one
$d_{10},...,d_{19}, d_{43}, d_{44}$. Explicitly we obtained
\beq
\begin{array} {lll}
g_1 = 360 & g_2 = -1440 & g_4 = 240 \\
g_{13} = 12 & g_{14} = 24 & g_{15} = 15 \\
g_{16} = -270 & g_{17} = 120 & g_{18} = 960 \\
g_{19} = 600 & g_{20} = -16 & g_{21} -17 \\
g_{22} = -10 & g_{25} = (60^-,195/2^+) & g_{27} = 90 \\
g_{28} = -270 & g_{29} = (450^-,630^+) & g_{33} = -90 \\
d_1 = (450^-,-90^+) & d_2 = (42^-,-111/2^+) & d_3 = (0^-,30^+) \\
d_4 = 240 & d_5 = 420 & d_6 = 390 \\
d_7 = 480 & d_8 = 420 & d_9 = 60 \\
d_{20} = 30 & d_{21} = -60 & d_{22} = -180 
\end{array} 
\eeq
The first group is completely determined using the relations
\beq
\begin{array}{ll}
\underline{\mbox{Term}} & \underline{\mbox{Coefficient}} \\
F_{;mm} K_{ab} K^{ab} & 0= -(m-2) g_1 +4(m-1) g_6 +4(m-1) g_{13} 
          +2g_{15} +d_3 +\frac 1 2 (m-2) g_{41} \\
                & \phantom{0=}  -2(m-1) g_{43} -(m-1) d_{24}
          -d_{27} +d_{28} -(m-5) d_{54} \\
F_{;mm} K^2 & 0= -2(m-1) g_6 +\frac 1 2 (m-2) d_1 -2(m-1) d_2 +\frac 1 2
              (m-2) g_{40} -2(m-1) g_{42} \\
                & \phantom{0=} -(m-1) d_{23} -d_{25} +d_{26} 
                -(m-5) d_{53} \\
F_{;m} KR & 0= \frac 1 2 (m-2) g_5 -2g_6 -4(m-1) g_7 -2g_8 -g_{11} -2d_2 -\frac
        1 2 (m-2) g_{39} \\
        & \phantom{0=} +2(m-1) g_{42} +2g_{43} +d_{25} -(m-5) g_{45} \\
F_{;m} KR_{mm} & 0= \frac 1 2 (m-2) g_1 +\frac 1 2 (m-2) g_{10} 
         -2(m-1) g_{11} -2(m-1) g_{13} +4g_{15} +g_{20} \\
           & \phantom{0=}  -2g_{21} -\frac 1 2
         (m-2) d_1 +2(m-1) d_2 +d_3 -\frac 1 2 (m-2) d_{20} +2(m-1) d_{23}\\
            & \phantom{0=} 
           +2d_{24} -d_{25} -d_{26} -(m-5) d_{41} \\
F_{;m} K_{ab} R^{ab} & 0= -\frac 1 2 (m-2) g_1 +2(m-1) g_6 -2(m-2) g_8 -8g_9 
        +2(m-1) g_{13} -4g_{15} +g_{20} \\
        & \phantom{0=}  -d_3 -\frac 1 2 (m-2) d_{21} 
          +(m-1) d_{25} +2d_{27} +2d_{29} -(m-5) d_{45} \\
F_{;m} K^{ab} R_{ammb} & 0= -(m-2) g_1 +4(m-1) g_6 +4(m-1) g_{13} 
          +2g_{15} -(m-2) g_{20} +2g_{22} -d_3\\
        & \phantom{0=}  -\frac 1 2 (m-2) d_{22} 
          +(m-1) d_{26} +2d_{28} +2d_{29} -(m-5) d_{46} \\
F_{;m} K_{ab} K^{bc} K_c^a & 0= (m-2) g_1 -4(m-1) g_6 -4(m-1) g_{13} 
          -2g_{15} -d_3 -(m-2) d_{27} +d_{28} \\
             & \phantom{0=} +2d_{29} 
          -\frac 1 2 (m-2) d_{35} +(m-1) d_{39} +4d_{40} -(m-5) d_{51} \\ 
FR_{ac} K^c_b K^{ab} & 0= -2(m-2) g_8 -8g_9 +4g_{15} +g_{20} +2d_3 
           +4d_{13} +4d_{14} -4d_{19} -(m-2) d_{27} +d_{28} +2d_{29} 
\end{array}
\eeq
One finds
\beq
\begin{array}{lll}
d_{23} = (-215/16^-,-275/16^+) & d_{24} = (-215/8 ^-, -275/8^+) &
d_{25} = (14^-,-1^+) \\
d_{26} = (-49/4^-, -109/4^+) & d_{27} = 16 & 
d_{28} = (47/2^-,-133/2^+) \\
d_{29} = 32 & d_{38} = (777/32^-, 375/32^+) & d_{39} = (17/2^-, 25^+) \\
d_{41} = (-255/8^-,165/8^+) & d_{45} =(-30^-,-15^+) & 
d_{46} = (-465/4^-,-165/4^+) 
\end{array}
\eeq
Finally let us consider the second group mentioned above. 
As we will see, one needs just one more relation in addition to those
obtained from equation (\ref{8}), which are, in detail,
\beq
\begin{array}{ll}
\underline{\mbox{Term}} & \underline{\mbox{Coefficient}}  \\
FK_{:b}K^{:b} & 0=2(m-1) g_6 -4g_{15} -(m-2) g_{20} +2g_{22} 
             - \frac 1 2 (m-2) d_1 +2(m-1) d_2 +2d_{10} \\
              & \phantom{0=} +d_{11} 
               -(m-3) d_{15} -d_{16} -d_{18} +(m-2) g_{40} 
              -4(m-1) g_{42} -2d_{23} -2d_{25} \\
FK_{ab:}^{\phantom{ab:}a} K_:^b & 0=2(m-2) g_1 -4(m-1) g_6 -8(m-1) g_{13} 
                +(4m-6) g_{20} -8g_{22} -(m-2) d_1 \\
         & \phantom{0=}+4(m-1) d_2 -(m-3) d_{11} 
                +2d_{12} -2d_{14} +2d_{16} -2d_{17} +2d_{18} -2(m-2) d_{25}
                +2d_{26} -4d_{29}  \\
FK_{ab:c} K^{ab\phantom{:}c}_{\phantom{ab}:} & 0=(m-2) g_1 -4(m-1) g_6 
              - 4(m-1) g_{13} -2g_{15} +(m-2) g_{20} -2g_{22} -3d_3 \\
     & \phantom{0=}         +2d_{13} +2d_{14} -(m-3) d_{19} +(m-2) g_{41} 
               -4(m-1) g_{43} -2d_{24} -2d_{27}  \\
FKK_{ab:}^{\phantom{ab:}ab} & 0= 4(m-2) g_8 +16g_9 -4g _{15} -mg_{20} +2g_{22}
                +d_{11} +2d_{12} -(m-4) d_{16}\\
          & \phantom{0=} -2d_{17} -d_{18} -(m-2) d_{25} 
                +d_{26} -2d_{29} \\
F_{;m} K_{:a}^{\phantom{:a} a} & 0= -\frac 3 2 (m-2) g_1 +4(m-1) g_6 -4(m-2) g_8
                -16g_9 +6(m-1) g_{13} +\frac 1 2 (m-2) d_1 \\
          & \phantom{0=}-2(m-1) d_2 -d_3 
                -\frac 1 2 (m-2) d_4 +\frac 1 2 (m-2) d_6 -\frac 1 2 (m-2) 
                 d_8 -2(m-1) d_{10}\\
        & \phantom{0=}  -d_{11} -2d_{13} +2(m-1) d_{15} 
                 +d_{16} +d_{18} +2d_{19} -(m-5) d_{43} \\
F_{;m} K_{ab:}^{\phantom{ab:}ab} & 0= \frac 1 2 (m-2) g_1 -2(m-1) g_6 
             +4(m-2) g_8 +16g_9 -2(m-1) g_{13} +2g_{15} +2d_3\\
        & \phantom{0=} -\frac 1 2 
            (m-2) d_5 +\frac 1 2 (m-2) d_7 -\frac 1 2 (m-2) d_9 
            -(m-1) d_{11} -2d_{12} -2d_{14}\\
         & \phantom{0=} +(m-1) d_{16} +2d_{17} +
               (m-1) d_{18} -(m-5) d_{44} \\
FK_{ab:}^{\phantom{ab:} a} K^{bc}_{\phantom{bc}:c} & 0= (4-3m) g_{20} 
                +6g_{22} -2d_3 -2(m-2) d_{12} -4d_{13} -2d_{14} \\
         & \phantom{0=} 
          +(m+1) d_{17} +4d_{19} -(m-2) d_{27} +d_{28} +2d_{29} \\
F K_{:ab} K^{ab} & 0= 2(m-2) g_1 -4(m-1) g_6 -2(m-2) g_8 -8g_9 
             -8 (m-1) g_{13} +(4m-5) g_{20}\\
          & \phantom{0=} -8g_{22} -(m-2) d_1 
             +4(m-1) d_2 -(m-2) d_{11} -2d_{14} +(m-2) d_{16} 
            +3d_{18}\\
       & \phantom{0=} -(m-2) d_{25} +d_{26} -2d_{29} 
\end{array}
\eeq
This yields 
\beq
\begin{array}{ll}
d_{11} = (58^-,238^+) & d_{15} = (6^-, 111^+) \\
d_{16} = (-30^-,-15^+) & d_{19} = (54^-,114^+) 
\end{array} 
\eeq
together with the relations
\beq
\begin{array}{ll}
2d_{10} +d_{43} = -91 & 2d_{10} -d_{18} = (-983/8^-,-1403/8^+) \\
2d_{14} -3d_{18} = (-913/4^-,-2533/4^+) & d_{13} +d_{14} = (297/8^-,837/8^+) \\
d_{18}-d_{44} = (60^-,225^+) & 2d_{12} -2d_{17} -d_{18} = (-7/4^-,-787/4^+) \\
2d_{12} -d_{17} = 32 & 
\end{array}
\eeq
This is all we can get with the relation (\ref{8}). It is seen, that 
for example given $d_{43}$ or $d_{44}$ the remaining constants may be 
determined. This is achieved with the equation (\ref{9}) \cite{17}.
Thus at the end one gets
\beq
\begin{array}{lll}
d_{10} =(-413/16^-,487/16^+) & d_{12} = (-11/4^-,49/4^+) & 
d_{13} = (355/8^-,535/8^+) \\
d_{14} = (-29/4^-,151/4^+) & d_{17} = (-75/2^-,-15/2^+) & 
d_{18} = (285/4^-,945/4^+) \\
d_{43} = (-315/8^-,-1215/8^+) & d_{44} = 45/4 & 
\end{array}
\eeq
In summary, we have determined the full $a_5$ heat-kernel coefficient
for Dirichlet and Robin boundary conditions.
All terms not displayed in the above lists have been used as a check 
for the universal constants. 

We have shown that special case evaluation of heat kernel coefficients 
supplemented by the conformal techniques developed in \cite{8} 
provide a very powerful tool for the calculation of the coefficients 
on general manifolds. The techniques displayed might prove very 
useful in finding universal constants for recently discussed generalized 
boundary conditions \cite{tang1,tang2,tang3,tang4,tang5,tang6} having 
relevance for one-loop quantum gravity and gauge theory.   
However, in these cases the task will be even more difficult due to the many
additional possibilities of building geometrical invariants 
\cite{tang3,tang6}. 

\acknowledgements
I'm indebted with Peter Gilkey for providing results on conformal 
rescalings. They have served as a very good check of the calculation and 
have been of invaluable help. Furthermore I wish to thank Stuart Dowker,
Giampiero Esposito and Michael Bordag for enlightening discussions as 
well as for help. 

This investigation has been partly supported by the DFG under contract 
number BO1112/4-2.

\end{document}